\documentclass[11pt]{article}
\usepackage{amssymb,amsmath,amsfonts}
\usepackage{graphicx}
\usepackage{graphics}
\usepackage{eepic,epsfig}


\begin{document}

\title{Radiation of surface polaritons by an annular beam coaxially
enclosing a cylindrical waveguide}
\author{A.A. Saharian$^{1,2}$, G.V. Chalyan$^{2}$, L.Sh. Grigoryan$^{1}$,
H.F. Khachatryan$^{1}$ \and V.Kh. Kotanjyan$^{1,2}$, \vspace{0.3cm} \\
\textit{$^1$Institute of Applied Problems of Physics NAS RA, }\\
\textit{25 Hr. Nersessian Str., 0014 Yerevan, Armenia} \vspace{0.3cm}\\
\textit{$^2$Institute of Physics, Yerevan State University,}\\
\textit{1 Alex Manoogian Street, 0025 Yerevan, Armenia }}
\maketitle

\begin{abstract}
We investigate the radiation of surface polaritons by an annular beam that
coaxially encloses a cylindrical waveguide surrounded by a homogeneous
medium. By using the Green dyadic, the electromagnetic potentials and the
electric and magnetic fields are found inside and outside the waveguide. The
expression for the energy losses is derived for the general case of the
dispersion for dielectric permittivities inside and outside the cylinder. A
comprehensive analysis is presented in the spectral range corresponding to
the radiation of surface polaritons. The highest peaks in the spectral
distribution are obtained for intermediate values of the beam velocity. In
the limit of transparent medium the spectrum of radiated surface polaritons
is discrete and the corresponding frequencies are determined by the
eigenvalue equation for the cylindrical waveguide. Numerical examples are
presented for the Drude model of dispersion.
\end{abstract}

\bigskip


\bigskip

\section{Introduction}

Surface polaritons are a class of surface waves that occur at the interface
between two media when the real parts of their dielectric permittivities
have opposite signs \cite{Maie07}-\cite{Stoc18}. They are collective
excitations of the electronic subsystem and the electromagnetic field
localized in a thin surface layer. The interest in surface polaritons is due
to their important properties, such as relatively high electromagnetic
energy densities, high sensitivity, and subwavelength resolution. However, a
significant challenge in their practical applications is the substantial
absorption in the corresponding spectral range. A pivotal research direction
involves the development of materials and metamaterials that exhibit reduced
absorption of surface polaritons within the desired frequency range (see,
for example, \cite{Marq08}-\cite{Qin23}).

Another important point related to the physics of surface polaritons is the
development of efficient mechanisms for their generation. Currently used
methods (see, e.g., \cite{Maie07,Enoc12,Han13,Abaj10}) include prism and
grating coupling to free space electromagnetic waves, coupling to guided
modes of waveguide, tight-focus and near-field scattering excitations.
Another class of mechanisms is based on the interaction of a beam of charged
particles with the interface around of which the surface polaritons are
located. The geometries of beams parallel and perpendicular to a planar
boundary have been discussed in the literature \cite{Bash06}-\cite{Zhan24}.
In particular, beams of scanning and transmission electron microscopes can
be used as sources of generation. The momentum of the electrons in the beam
is essentially greater in comparison to that of photons. This disparity
enables the excitation of surface polaritons with greater ease under a
relatively wide range of conditions, obviating the necessity for coupling
elements such as prisms or gratings. The availability of highly focused
electron beams in both space and time facilitates precise control over the
excitation of surface polaritons at specific locations. This capability
confers an important advantage for the selective generation of plasmonic
modes in nanostructures. Another important application of beam-induced
generation of surface polaritons is electron energy loss spectroscopy, a
crucial tool in electron microscopy that provides detailed information about
the plasmonic properties of materials and serves as a tool for beam
diagnostics \cite{Abaj10,Jian17}. In using this class of mechanisms it
should be taken into account that other types of electromagnetic radiation,
for example, Cherenkov, diffraction and transition radiations, may be
excited. The total energy losses for planar, spherical, and cylindrical
boundaries have been studied in the literature (see, e.g., \cite%
{Abaj10,Riva00}). More complicated structured geometries were considered as
well.

In \cite{Kota18}-\cite{Saha23} the generation of surface polaritons by a
charged particle is investigated on a cylindrical interface between two
media with different dielectric permittivities. A single particle moving
parallel to the axis of the dielectric cylinder and circulating around that
axis were considered. From the point of view of practical application, it is
important to generalize the obtained results for particle beams. In the
present paper we study the radiation of surface polaritons by an annular
beam coaxially enclosing the cylindrical waveguide. The hollow structure
enhances the efficiency of coupling between the beam and the electromagnetic
modes of the system. Various ~applications of annular beams in condensed
matter physics, materials science, and high-energy physics can be found in
the literature (see, for example, \cite{AnBe05,Lloy17,Liu24} and references
therein). These applications include the acceleration and collimation of
charged particles, X-ray generation, manipulation of nanoparticles, surface
treatment and deposition processes, and free-electron lasers. The present
study aims to demonstrate that the electron annular beams can serve as
sources of surface polaritons propagating along a cylindrical interface
between two media.

The paper is organized as follows. In the next section we describe the
geometry of the problem. The partial Fourier components of the
electromagnetic potentials and the electric and magnetic field strengths are
presented. In Section \ref{sec:Losses} the energy losses are studied for the
general case of dispersions for dielectric permittivities of the cylinder
and surrounding medium. The energy radiated in the form of surface
polaritons is discussed in Section \ref{sec:SPs}. The results of the
corresponding numerical evaluations are presented. The main results are
summarized in Section \ref{sec:Conc}.

\section{Problem setup and the electromagnetic field}

\label{sec:Fields}

The setup of the problem under consideration is illustrated in Fig. \ref%
{fig1}. A thin annular beam of charged particles moves coaxially outside a
cylinder of radius $r_{c}$ and with dielectric permittivity $\varepsilon
_{0} $. The general case of surrounding medium with dielectric permittivity $%
\varepsilon _{1}$ will be considered. The cylindrical coordinates $(r,\phi
,z)$ will be used with the axis $z$ along the cylinder axis. The current
density for the annular beam of radius $r_{0}$ is given by the expression 
\begin{equation}
j_{l}(x)=\delta _{3l}\frac{qv}{r}\delta (r-r_{0})\delta (z-vt),  \label{jl}
\end{equation}%
where $v$ is the velocity of the charges and $\delta _{kl}$ is the Kronecker
symbol. Here and below $x$ is used to denote the spacetime point $%
x=(t,r,\phi ,z)$ and the indices $i,l=1,2,3$ correspond to the cylindrical
components $r,\phi ,z$ of the vectors. The charge density is expressed as $%
\rho (x)=q\delta (r-r_{0})\delta (z-vt)/r$ and for the total charge of the
beam one has $Q=2\pi q$. We are interested in the radiation of surface
polaritons propagating along the cylinder surface.

\begin{figure}[tbph]
\begin{center}
\epsfig{figure=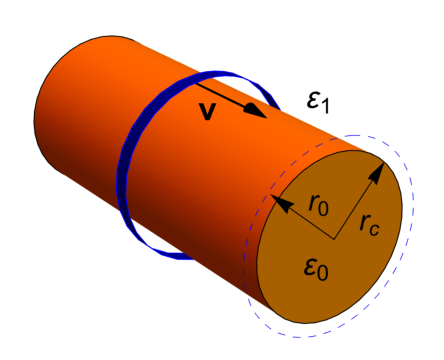,width=6cm,height=5cm}
\end{center}
\caption{Setup of the problem.}
\label{fig1}
\end{figure}

The electromagnetic fields are found by using the Green dyadic $%
G_{il}(x,x^{\prime })$. In the Lorentz gauge the components of the vector
potential $\mathbf{A}(x)$ are expressed as 
\begin{equation}
A_{i}(x)=-\int d^{4}x^{\prime }\sum_{l=1}^{3}\frac{G_{il}(x,x^{\prime })}{%
2\pi ^{2}c}j_{l}(x^{\prime }).  \label{AiG}
\end{equation}%
The geometry of the problem under consideration is homogeneous with respect
to the spacetime coordinates $t$, $\phi $, $z$. From here it follows that
the Green dyadic depends on the corresponding arguments in the form $%
t-t^{\prime }$, $\phi -\phi ^{\prime }$, and $z-z^{\prime }$. By taking into
account that the problem is periodic with respect to the coordinate $\phi $,
we use the partial Fourier expansion 
\begin{align}
G_{il}(x,x^{\prime })& =\sum_{n=-\infty }^{\infty }\int_{-\infty }^{\infty
}d\omega \int_{-\infty }^{\infty }dk_{z}\,G_{il,n}(\omega ,k_{z},r,r^{\prime
})  \notag \\
& \times e^{in(\phi -\phi ^{\prime })+ik_{z}(z-z^{\prime })-i\omega
(t-t^{\prime })}.  \label{Gexp}
\end{align}%
Substituting this expansion and the expression (\ref{jl}) in (\ref{AiG}),
the integrals over $r^{\prime }$ and $z^{\prime }$ are evaluated with the
help of delta functions. The integral over $\phi ^{\prime }$ gives $2\pi
\delta _{0n}$ and for the integral over $t^{\prime }$ we get $2\pi \delta
(\omega -k_{z}v)$. The integration over $\omega $ is done by using the delta
function $\delta (\omega -k_{z}v)$ and this gives $\omega =k_{z}v$. In this
way, for the Fourier components of the vector potential we get 
\begin{equation}
A_{lk_{z}}(r)=-2q\beta G_{l3,0}(k_{z}v,k_{z},r,r_{0}),  \label{AiF}
\end{equation}%
where $\beta =v/c$. Here and below the Fourier component $f_{k_{z}}(r)$ of
the field $f(x)$ is defined by the relation 
\begin{equation}
f(x)=\int_{-\infty }^{\infty }dk_{z}\,f_{k_{z}}(r)e^{ik_{z}\left(
z-vt\right) }=2\mathrm{Re}\left[ \int_{0}^{\infty
}dk_{z}\,f_{k_{z}}(r)e^{ik_{z}\left( z-vt\right) }\right] ,  \label{fexp}
\end{equation}%
where the relation $f_{-k_{z}}(r)=f_{k_{z}}^{\ast }(r)$ is used, valid for a
real function $f(x)$. The problem under consideration is azimuthally
symmetric and the only nonzero contribution to the fields comes from the
mode $n=0$. The second representation in (\ref{fexp}) shows that without
loss of generality we can assume that $k_{z}\geq 0$.

In \cite{Grig95} a general scheme is developed for the construction of the
Green dyadic in cylindrically symmetric piecewise homogeneous media. By
using the corresponding expressions for $G_{il,n}(\omega ,k_{z},r,r^{\prime
})$ in (\ref{AiF}), in the region outside the cylinder, $r>r_{c}$, the
nonzero components of the vector potential are presented as%
\begin{align}
A_{1k_{z}}(r)& =-2iq\beta D(k_{z})H_{0}(u_{1}\frac{r_{0}}{r_{c}})H_{1}(u_{1}%
\frac{r}{r_{c}}),  \notag \\
A_{3k_{z}}(r)& =i\pi q\beta \left[ J_{0}(u_{1}\frac{r_{<}}{r_{0}})H_{0}(u_{1}%
\frac{r_{>}}{r_{0}})-\frac{U_{0}^{J}}{U_{0}^{H}}H_{0}(u_{1}\frac{r_{0}}{r_{c}%
})H_{0}(u_{1}\frac{r}{r_{c}})\right] ,  \label{A13e}
\end{align}%
where $r_{<}=\mathrm{min}\,(r_{0},r)$, $r_{>}=\mathrm{max}\,(r_{0},r)$, $%
J_{n}(y)$ and $H_{n}(y)=H_{n}^{(1)}(y)$ are the Bessel and Hankel functions,
and%
\begin{equation}
u_{j}=u\left( \beta ^{2}\varepsilon _{j}-1\right) ^{\frac{1}{2}%
},\;u=k_{z}r_{c},  \label{uj}
\end{equation}%
for $j=0,1$. The function $D(k_{z})$ is defined by 
\begin{equation}
D(k_{z})=\left( \varepsilon _{0}-\varepsilon _{1}\right) \frac{uJ_{0}(u_{0})%
}{U_{0}^{H}U(u)}\left\{ 
\begin{array}{cc}
H_{1}(u_{1}), & r<r_{c} \\ 
J_{1}(u_{0}), & r>r_{c}%
\end{array}%
\right. ,  \label{Dk}
\end{equation}%
with the function%
\begin{equation}
U(u)=\varepsilon _{1}u_{0}J_{0}(u_{0})H_{1}(u_{1})-\varepsilon
_{0}u_{1}J_{1}(u_{0})H_{0}(u_{1}).  \label{Uu}
\end{equation}%
Here and below we use the notation 
\begin{align}
U_{n}^{F}& =u_{1}J_{n}(u_{0})F_{n}^{\prime
}(u_{1})-u_{0}F_{n}(u_{1})J_{n}^{\prime }(u_{0})  \notag \\
& =-u_{1-n}J_{0}(u_{0})F_{1}(u_{1})+u_{n}J_{1}(u_{0})F_{0}(u_{1}),
\label{Vn}
\end{align}%
for $F=J,H$, and $n=0,1$. In the region inside the cylinder, $r<r_{c}$, the
expressions for the nonzero components of the vector potential read%
\begin{align}
A_{1k_{z}}(r)& =-2iq\beta D(k_{z})H_{0}(u_{1}\frac{r_{0}}{r_{c}})J_{1}(u_{0}%
\frac{r}{r_{c}}),  \notag \\
A_{3k_{z}}(r)& =-\frac{2q\beta }{U_{0}^{H}}H_{0}(u_{1}\frac{r_{0}}{r_{c}}%
)J_{0}(u_{0}\frac{r}{r_{c}}).  \label{A13i}
\end{align}%
The formulas above are valid for all values {}{}of beam velocity. For large
values of the coordinate $r$ the radial dependence of the fields $%
A_{lk_{z}}(r)$ from (\ref{A13e}) is given by $r^{-1/2}\exp [ik_{z}\left(
\beta ^{2}\varepsilon _{1}-1\right) ^{\frac{1}{2}}r]$. This shows that, in
order to escape the exponential increase of the fields at infinity, for
complex values of the radical $\left( \beta ^{2}\varepsilon _{1}-1\right) ^{%
\frac{1}{2}}$ its sign should be taken in accordance with the condition $%
\mathrm{Im}[\left( \beta ^{2}\varepsilon _{1}-1\right) ^{\frac{1}{2}}]\geq 0$
for $k_{z}\geq 0$. With this choice the Hankel functions in (\ref{A13e}) are
expressed in terms of the modified Bessel functions $K_{n}(u\left( 1-\beta
^{2}\varepsilon _{1}\right) ^{1/2}r/r_{c})$, $n=0,1$, with positive real
part of the argument. In particular, this is the case for real $\varepsilon
_{1}$ and for the velocities in the range $\beta ^{2}\varepsilon _{1}<1$. As
for the choice of the sign of the radical in $u_{0}$, it enters in the
arguments of the Bessel functions $J_{n}(w)$, $n=0,1$, and both signs of the
root lead to the same expressions for the Fourier components.

The scalar potential $\varphi (x)$ is found from the gauge condition $%
(\varepsilon /c)\partial \varphi /\partial t$ $+\nabla \cdot \mathbf{A}=0$.
The expression for the Fourier component takes the form%
\begin{align}
\varphi _{k_{z}}(r)& =\frac{i\pi q}{\varepsilon _{1}}\left[ J_{0}(u_{1}\frac{%
r_{<}}{r_{c}})H_{0}(u_{1}\frac{r_{>}}{r_{c}})+H_{0}(u_{1}\frac{r_{0}}{r_{c}}%
)\right.  \notag \\
& \times \left. H_{0}(u_{1}\frac{r}{r_{c}})\left( \frac{2iu_{1}}{\pi u}%
D(k_{z})-\frac{U_{0}^{J}}{U_{0}^{H}}\right) \right] ,  \label{phie}
\end{align}%
in the region $r>r_{c}$ and 
\begin{equation}
\varphi _{k_{z}}(r)=-\frac{2q}{\varepsilon _{0}}H_{0}(u_{1}\frac{r_{0}}{r_{c}%
})J_{0}(u_{0}\frac{r}{r_{c}})\left( \frac{u_{0}}{u}D(k_{z})+\frac{1}{%
U_{0}^{H}}\right) ,  \label{phii}
\end{equation}%
for $r<r_{c}$. The Fourier components have poles at the zeros of the
function $U(u)$. Those poles correspond to the eigenmodes of the cylinder
with respect to $u=k_{z}r_{c}$ for given $\beta $ and $\varepsilon _{j}$.

The electric field is obtained by using the relation $\mathbf{E}%
=-(1/c)\partial \mathbf{A}/\partial t-\mathbf{\nabla }\,\varphi $. For the
nonzero Fourier components this gives%
\begin{align}
E_{1k_{z}}(r)& =\frac{\pi q}{i\varepsilon _{1}r_{c}}\left[ r_{c}\frac{%
\partial }{\partial r}J_{0}(u_{1}\frac{r_{<}}{r_{c}})H_{0}(u_{1}\frac{r_{>}}{%
r_{c}})\right.  \notag \\
& +u_{1}H_{0}(u_{1}\frac{r_{0}}{r_{c}})H_{1}(u_{1}\frac{r}{r_{c}})\left.
\left( \frac{U_{0}^{J}}{U_{0}^{H}}+\frac{2iu}{\pi u_{1}}D(k_{z})\right) %
\right] ,  \notag \\
E_{3k_{z}}(r)& =\frac{\pi qk_{z}}{\varepsilon _{1}}\left( 1-\beta
^{2}\varepsilon _{1}\right) \left[ J_{0}(u_{1}\frac{r_{<}}{r_{c}})H_{0}(u_{1}%
\frac{r_{>}}{r_{c}})\right.  \notag \\
& -\left. \left( \frac{U_{0}^{J}}{U_{0}^{H}}+\frac{2iu}{\pi u_{1}}%
D(k_{z})\right) H_{0}(u_{1}\frac{r_{0}}{r_{c}})H_{0}(u_{1}\frac{r}{r_{c}})%
\right] ,  \label{E13ex}
\end{align}%
outside the cylinder, $r>r_{c}$, and 
\begin{equation}
E_{lk_{z}}(r)=\frac{2q}{\varepsilon _{0}r_{c}}H_{0}(u_{1}\frac{r_{0}}{r_{c}}%
)\left( D(k_{z})-\frac{u_{0}}{uU_{0}^{H}}\right) \left\{ 
\begin{array}{cc}
uJ_{1}(u_{0}\frac{r}{r_{c}}), & l=1 \\ 
iu_{0}J_{0}(u_{0}\frac{r}{r_{c}}), & l=3%
\end{array}%
\right. ,  \label{E13i}
\end{equation}%
in the interior region, $r<r_{c}$. For the magnetic field we have $\mathbf{H}%
=\nabla \times \mathbf{A}$. By using the expressions for the vector
potential, for the Fourier components of the magnetic field it can be seen
that $H_{1k_{z}}(r)=H_{3k_{z}}(r)=0$ and 
\begin{equation}
H_{2k_{z}}(r)=\beta \varepsilon E_{1k_{z}}(r),\;\varepsilon =\varepsilon
_{0}\theta (r_{c}-r)+\varepsilon _{1}\theta (r-r_{c}),  \label{H2ei}
\end{equation}%
where $\theta (x)$ is the Heaviside unit step function. As seen, the
electric and magnetic fields are orthogonal and the magnetic field is
perpendicular to the cylinder axis (TM waves). The only nonzero component on
the cylinder axis corresponds to $E_{3k_{z}}(r)$.

\section{Energy losses}

\label{sec:Losses}

Having evaluated the electromagnetic fields we turn to the energy losses by
the annular beam. The work done by the field per unit length of the beam
trajectory is given by $dW/dz=QE_{3}(x)|_{r=r_{0},z=vt}$. The spectral
density of the energy losses per unit time, denoted here by $d\mathcal{E}%
/d\omega $, is related to the work done by the field through the formula 
\begin{equation}
\frac{dW}{dz}=-\frac{1}{v}\int_{0}^{\infty }d\omega \,\frac{d\mathcal{E}}{%
d\omega }.  \label{dWE}
\end{equation}%
Plugging the Fourier expansion (\ref{fexp}) for $E_{3}(x)$ and passing from
the integration over $k_{z}$ to the integration over $\omega =k_{z}v$, we get%
\begin{align}
\frac{d\mathcal{E}}{d\omega }& =\frac{d\mathcal{E}_{\mathrm{h}}}{d\omega }-%
\frac{Q^{2}}{c}\beta \omega \,\mathrm{Re\,}\Bigg\{\left( 1-\frac{1}{\beta
^{2}\varepsilon _{1}}\right) \frac{H_{0}^{2}(u_{1}\frac{r_{0}}{r_{c}})}{U(u)}
\notag \\
& \times \left[ \varepsilon _{1}u_{0}J_{0}(u_{0})J_{1}(u_{1})-\varepsilon
_{0}u_{1}J_{1}(u_{0})J_{0}(u_{1})\right] \Bigg\},  \label{dE}
\end{align}%
where 
\begin{equation}
\frac{d\mathcal{E}_{\mathrm{h}}}{d\omega }=\frac{Q^{2}}{c}\beta \omega \,%
\mathrm{Re}\left[ \left( 1-\frac{1}{\beta ^{2}\varepsilon _{1}}\right)
J_{0}(u_{1}\frac{r_{0}}{r_{c}})H_{0}(u_{1}\frac{r_{0}}{r_{c}})\right] ,
\label{dEh}
\end{equation}%
is the spectral density of the energy losses in a homogeneous medium with
permittivity $\varepsilon _{1}$. In these expressions $u_{0}$ and $u_{1}$
are given by (\ref{uj}), where $k_{z}=\omega /v$ and $u=\omega r_{c}/v$.

For real dielectric permittivities $\varepsilon _{0}$ and $\varepsilon _{1}$
the possible channels of the energy losses are in the form of different
types of radiation processes. They correspond to the Cherenkov radiation in
the exterior medium under the condition $\beta \sqrt{\varepsilon _{1}}>1$
(for the features of the Cherenkov radiation by a point charge moving
paraxially inside and outside the cylinder see \cite{Saha20,Saha24b}), to
the radiation on guiding modes of the cylindrical waveguide under the
conditions $\beta \sqrt{\varepsilon _{1}}<1<\beta \sqrt{\varepsilon _{0}}$,
and to surface polaritons. The latter are radiated in the spectral range
where the dielectric permittivities $\varepsilon _{0}$ and $\varepsilon _{1}$
have opposite signs and the Cherenkov condition in the medium with positive
permittivity is not satisfied. The spectral density of the Cherenkov
radiation intensity in a homogeneous medium with permittivity $\varepsilon
_{1}$, in the spectral range $\beta ^{2}\varepsilon _{1}(\omega )>1$, is
presented in the form 
\begin{equation}
\frac{d\mathcal{E}_{\mathrm{h}}}{d\omega }=\frac{Q^{2}}{c}\beta \omega
\,\left( 1-\frac{1}{\beta ^{2}\varepsilon _{1}}\right) J_{0}^{2}\left( \frac{%
\omega }{c}r_{0}\sqrt{\varepsilon _{1}-\beta ^{-2}}\right) .  \label{dEh2}
\end{equation}%
In the limit $r_{0}\rightarrow 0$ this formula is reduced to the one for a
point charge $Q$. Note that the radiation intensity (\ref{dEh2}) becomes
zero for frequencies corresponding to the zeros of the Bessel function $%
J_{0}(x)$.

Our main interest here is the energy losses in the form of surface
polaritons. The details of the Cherenkov radiation in the exterior medium
and of the radiation on guiding modes of the cylinder will be discussed
elsewhere.

\section{Radiation of surface polaritons}

\label{sec:SPs}

For surface polaritons the real parts $\varepsilon _{0}^{\prime }$ and $%
\varepsilon _{1}^{\prime }$ of the permittivities $\varepsilon _{0}$ and $%
\varepsilon _{1}$ should have opposite signs. We will consider the case $%
\varepsilon _{0}^{\prime }<0<\varepsilon _{1}^{\prime }$ and $\beta
^{2}\varepsilon _{1}^{\prime }<1$. This indicates that the Cherenkov
condition ($\beta ^{2}\varepsilon _{j}^{\prime }>1$) is not met in either
the interior or exterior media. It is convenient to introduce $\gamma
_{j}=\left( 1-\beta ^{2}\varepsilon _{j}\right) ^{\frac{1}{2}}$ with $%
u_{j}=i\gamma _{j}u$, where $u=\omega r_{c}/v$. Introducing the modified
Bessel functions $I_{\nu }(x)$ and $K_{\nu }(x)$, and considering the
special case of real $\varepsilon _{1}$, the formula for the spectral
density of the energy losses is presented as 
\begin{align}
\frac{d\mathcal{E}}{d\omega }& =\frac{2Q^{2}}{\pi c}\beta \omega \left( 
\frac{1}{\beta ^{2}\varepsilon _{1}}-1\right) \,\mathrm{Im}\Bigg\{\frac{%
K_{0}^{2}(w_{1}\frac{r_{0}}{r_{c}})}{U_{\mathrm{sp}}(u)}  \notag \\
& \times \left[ \varepsilon _{0}w_{1}I_{1}(w_{0})I_{0}(w_{1})-\varepsilon
_{1}w_{0}I_{0}(w_{0})I_{1}(w_{1})\right] \Bigg\},  \label{dEsp}
\end{align}%
where $w_{j}=u\left( 1-\beta ^{2}\varepsilon _{j}\right) ^{\frac{1}{2}}$ and 
\begin{equation}
U_{\mathrm{sp}}(u)=\varepsilon _{1}w_{0}I_{0}(w_{0})K_{1}(w_{1})+\varepsilon
_{0}w_{1}I_{1}(w_{0})K_{0}(w_{1}).  \label{Usp}
\end{equation}%
Note that for real $\varepsilon _{1}$ and $\beta ^{2}\varepsilon _{1}<1$ we
have $d\mathcal{E}_{\mathrm{h}}/d\omega =0$. The result (\ref{dEsp}) for the
spectral density of the energy losses is valid for general case of the
dispersion law $\varepsilon _{j}=\varepsilon _{j}(\omega )$, $j=0,1$. The
dependence on the beam radius enters through the function $%
K_{0}^{2}(w_{1}r_{0}/r_{c})$ and the energy losses exponentially decay for $%
r_{0}>\lambda _{\mathrm{sp}}/(2\pi \gamma _{1})$, where $\lambda _{\mathrm{sp%
}}$ is the radiation wavelength. As it follows from the formulas in the
previous section, the electric and magnetic fields for surface polaritons
are orthogonal and TM waves are radiated.

Let us consider the behavior of the energy losses (\ref{dEsp}) for limiting
values of the parameter $\beta _{1}=\beta \sqrt{\varepsilon _{1}}$. In the
limit $\beta _{1}\rightarrow 1$ one has $w_{1}\rightarrow 0$. By using the
asymptotics of the modified Bessel functions for small argument, in the
leading order we get%
\begin{equation}
\frac{d\mathcal{E}}{d\omega }\approx \frac{Q^{2}\sqrt{\varepsilon _{1}}}{%
2\pi r_{c}}\left( \frac{\omega r_{c}}{c}\right) ^{3}\left[ \left( 1-\beta
_{1}^{2}\right) \ln \left( 1-\beta _{1}^{2}\right) \right] ^{2}\mathrm{Im}%
\left[ \frac{\varepsilon _{0}I_{1}(w_{0})}{\varepsilon _{1}w_{0}I_{0}(w_{0})}%
-\frac{1}{2}\right] ,  \label{dEsp1}
\end{equation}%
where $w_{0}\approx u\sqrt{1-\varepsilon _{0}/\varepsilon _{1}}$. Hence, $d%
\mathcal{E}/d\omega $ tends to zero for $\beta _{1}\rightarrow 1$. In the
nonrelativistic limit, $\beta \ll 1$, we have $w_{j}\approx \omega
r_{c}/v\gg 1$ and the arguments of the modified Bessel functions are large.
In the leading approximation the behavior of the energy losses is described
by 
\begin{equation}
\frac{d\mathcal{E}}{d\omega }\approx \frac{2Q^{2}v}{\pi r_{c}^{2}\omega }%
\frac{\varepsilon _{0}^{\prime \prime }e^{-2\omega (r_{0}-r_{c})/v}}{\left(
\varepsilon _{1}+\varepsilon _{0}^{\prime }\right) ^{2}+\varepsilon
_{0}^{\prime \prime 2}},  \label{dEsp2}
\end{equation}%
with $\varepsilon _{0}^{\prime }$ and $\varepsilon _{0}^{\prime \prime }$
being the real and imaginary parts of $\varepsilon _{0}$. As expected, one
has $d\mathcal{E}/d\omega \rightarrow 0$ in the limit $\beta \rightarrow 0$.
Therefore, for both small and large velocities, the radiation intensity
tends to zero. The highest peaks in the spectral distribution are obtained
for intermediate values of the beam velocity.

\begin{figure}[tbph]
\begin{center}
\epsfig{figure=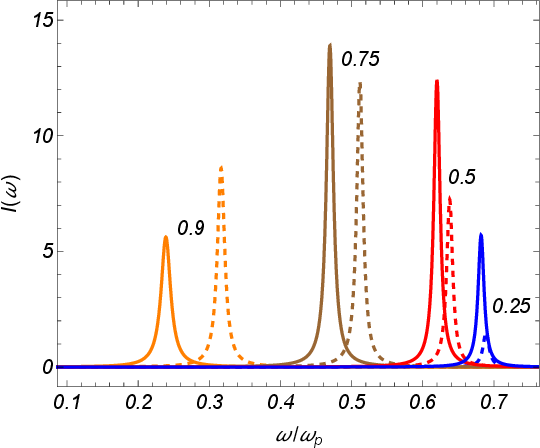,width=8cm,height=7cm}
\end{center}
\caption{Energy losses described by (\protect\ref{Iom}) versus the frequency
for $r_{0}/r_{c}=1.05$ and $\protect\gamma /\protect\omega _{p}=1/100$. The
full and dashed curves are plotted for $\protect\omega _{p}r_{c}/c=5$ and $%
\protect\omega _{p}r_{c}/c=10$ and the numbers near the curves are the
values of $\protect\beta $. It is assumed that the beam moves in the vacuum (%
$\protect\varepsilon _{1}=1$). }
\label{fig2}
\end{figure}

Here we will illustrate the results for the Drude model of the interior
dielectric permittivity, 
\begin{equation}
\varepsilon _{0}(\omega )=1-\frac{\omega _{p}^{2}}{\omega ^{2}+i\gamma
\omega },  \label{Drude1}
\end{equation}%
where $\omega _{p}$ is the plasma frequency and $\gamma $ is the damping
frequency. It will be assumed that the dispersion for the permittivity $%
\varepsilon _{1}$ is weak in the spectral range under consideration. In
particular, we can take $\varepsilon _{1}=1$, corresponding to the motion of
the beam outside a cylinder in the vacuum. The numerical results will be
presented for the dimensionless quantity 
\begin{equation}
I(\omega )=\frac{r_{c}}{Q^{2}}\frac{d\mathcal{E}}{d\omega }.  \label{Iom}
\end{equation}%
In Fig. \ref{fig2} the dependence of this quantity on the ratio $\omega
/\omega _{p}$ is displayed for different values of $\beta $ (the numbers
near the curves). The graphs are plotted for $\varepsilon _{1}=1$, $%
r_{0}/r_{c}=1.05$ and $\gamma /\omega _{p}=1/100$. The full and dashed
curves correspond to $\omega _{p}r_{c}/c=5$ and $\omega _{p}r_{c}/c=10$,
respectively.

\begin{figure}[tbph]
\begin{center}
\epsfig{figure=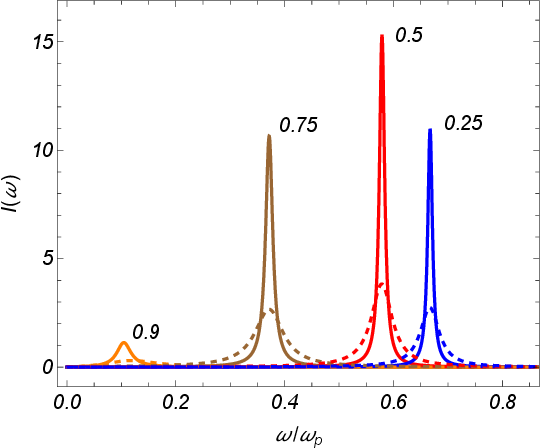,width=8cm,height=7cm}
\end{center}
\caption{The same as in Fig. \protect\ref{fig2} for $r_{0}/r_{c}=1.05$ and $%
\protect\omega _{p}r_{c}/c=2.5$. $\protect\gamma /\protect\omega _{p}=1/25$.
The full and dashed curves are plotted for $\protect\gamma /\protect\omega %
_{p}=1/100$ and $\protect\gamma /\protect\omega _{p}=1/25$.}
\label{fig3}
\end{figure}

To illustrate the dependence on the parameter $\gamma $ in (\ref{Drude1}),
in Fig. \ref{fig3} the quantity $I(\omega )$ is plotted versus $\omega
/\omega _{p}$ for $\gamma /\omega _{p}=1/100$ (full curves) and $\gamma
/\omega _{p}=1/25$ (dashed curves). For the other parameters we have taken
the values corresponding to the dimensionless combinations $\omega
_{p}r_{c}/c=2.5$ and $r_{0}/r_{c}=1.05$. As before, the numbers near the
curves are the values of $\beta $. As expected, the heights of the peaks
decrease with increasing $\gamma $ whereas their widths increase.

\begin{figure}[tbph]
\begin{center}
\epsfig{figure=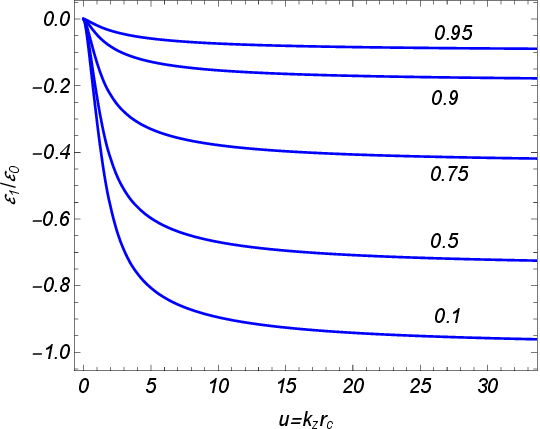,width=8cm,height=7cm}
\end{center}
\par
.
\caption{The solutions of the equation for surface polariton eigenmodes with
respect to the ratio $\protect\varepsilon _{1}/\protect\varepsilon _{0}$ as
functions of $u=k_{z}r_{c}$. The numbers near the curves correspond to the
values of $\protect\beta _{1}$.}
\label{fig4}
\end{figure}

In the idealized problem with real dielectric permittivities $\varepsilon
_{0}$ and $\varepsilon _{1}$ in the range $\varepsilon _{0}<0<\varepsilon
_{1}<1/\beta ^{2}$, the expression in the right-hand side of (\ref{dEsp})
has poles at the zeros of denominator. These poles correspond to the surface
polariton eigenmodes of the cylinder and the equation determining their
locations reads 
\begin{equation}
U_{\mathrm{sp}}(u)=0.  \label{SPmodes}
\end{equation}%
This equation is obtained from the dispersion equation for surface
polaritons in the general case of the azimuthal number $n$ (see, e.g., \cite%
{Kota19,Saha23,Ashl74,Khos91}) in the special case $n=0$ by using the
relations $I_{0}^{\prime }(w_{0})=I_{1}(w_{0})$ and $K_{0}^{\prime
}(w_{1})=-K_{1}(w_{1})$. The roots of equation (\ref{SPmodes}) depend on the problem
parameters in the form of two combinations $\beta _{1}=\beta \sqrt{%
\varepsilon _{1}}$ and $\varepsilon _{1}/\varepsilon _{0}$. It can be shown
that for a given $\beta _{1}$ the equation has a single root in the range%
\begin{equation}
\beta _{1}^{2}-1<\frac{\varepsilon _{1}}{\varepsilon _{0}}<0,
\label{regroot}
\end{equation}%
and there are no roots outside that range. We will denote the root by $u=u_{%
\mathrm{sp}}(\beta _{1},\varepsilon _{1}/\varepsilon _{0})$. For fixed $%
\beta _{1}$ the root $u_{\mathrm{sp}}$ is a monotonically decreasing
function of the ratio $\varepsilon _{1}/\varepsilon _{0}$. We have $u_{%
\mathrm{sp}}\rightarrow 0$ for $\varepsilon _{1}/\varepsilon _{0}\rightarrow
0$ and $u_{\mathrm{sp}}\rightarrow \infty $ for $\varepsilon
_{1}/\varepsilon _{0}\rightarrow \beta _{1}^{2}-1$. To see the values of the
ratio $\varepsilon _{1}/\varepsilon _{0}$ needed to have a radiation on a
given wavelength (determined by $u=k_{z}r_{c}$), we can consider (\ref%
{SPmodes}) as an equation with respect to $\varepsilon _{1}/\varepsilon _{0}$
for a given $u$. The dependence of the corresponding roots, as functions of $%
u$, is depicted in Fig. \ref{fig4} for different values of $\beta _{1}$
(numbers near the curves). With increasing velocity of the beam the range of
the ratio $\varepsilon _{1}/\varepsilon _{0}$ allowing the existence of
surface polaritons becomes narrower.

For real $\varepsilon _{0}$ and $\varepsilon _{1}$ in the range $\varepsilon
_{0}<0<\varepsilon _{1}<1/\beta ^{2}$ the only energy losses are in the form
of surface polaritons. In this case the expression under the imaginary part
sign in (\ref{dEsp}) is real and the nonzero contribution to the total
energy losses $\mathcal{E}_{\omega _{\mathrm{sp}}}=\int d\omega \,(d\mathcal{%
E}/d\omega )$ comes from the pole $\omega =\omega _{\mathrm{sp}}=vu_{\mathrm{%
sp}}/r_{c}$ with $u_{\mathrm{sp}}$ being the root of the equation (\ref%
{SPmodes}). In order to specify the integration contour near the pole we
introduce a small imaginary part of the permittivity $\varepsilon _{0}$
writing it in the form $\varepsilon _{0}=\varepsilon _{0}^{\prime
}+i\varepsilon _{0}^{\prime \prime }$. Expanding with respect to the small
imaginary part $\varepsilon _{0}^{\prime \prime }$, we get%
\begin{align}
\frac{d\mathcal{E}}{d\omega }& =-Q^{2}\frac{2\omega \gamma _{1}^{2}}{\pi
v\varepsilon _{1}}\left[ \frac{\varepsilon _{1}w_{0}I_{0}(w_{0})I_{1}(w_{1})%
}{\varepsilon _{0}w_{1}I_{1}(w_{0})I_{0}(w_{1})}-1\right]  \notag \\
& \times \mathrm{Im}\left[ \frac{\frac{I_{0}(w_{1})}{K_{0}(w_{1})}%
K_{0}^{2}(w_{1}\frac{r_{0}}{r_{c}})}{U_{\mathrm{sp}}(u)+\frac{i\varepsilon
_{0}^{\prime \prime }}{\varepsilon _{0}}B}\right] _{\varepsilon
_{0}=\varepsilon _{0}^{\prime }}.  \label{dE1}
\end{align}%
where 
\begin{equation}
B=1+\frac{\varepsilon _{0}\beta ^{2}}{2\gamma _{0}}u\left[ \frac{I_{1}(w_{0})%
}{I_{0}(w_{0})}-\frac{I_{0}(w_{0})}{I_{1}(w_{0})}\right] .  \label{B}
\end{equation}%
By taking into account that $0\leq I_{1}(x)/I_{0}(x)<1$ for $x\geq 0$ and $%
\varepsilon _{0}^{\prime \prime }(\omega )>0$ for $\omega >0$, we conclude
that in the spectral range with $\varepsilon _{0}^{\prime }<0$ one has $B>0$%
. The limit $\varepsilon _{0}^{\prime \prime }\rightarrow 0$ is taken by
using the formula 
\begin{equation}
\lim_{a\rightarrow 0+}\mathrm{Im}\left( \frac{1}{x-ia}\right)
=\lim_{a\rightarrow 0+}\frac{a}{x^{2}+a^{2}}=\pi \delta (x).
\label{deltalim}
\end{equation}%
By making use of this relation in (\ref{dE1}) and integrating over
frequency, for the energy radiated per unit time on a given surface
polariton mode $u=u_{\mathrm{sp}}$ we find%
\begin{equation}
\mathcal{E}_{\omega _{\mathrm{sp}}}=\frac{2Q^{2}\omega _{\mathrm{sp}}}{%
r_{c}\varepsilon _{1}}\frac{K_{0}^{2}(w_{1}\frac{r_{0}}{r_{c}})}{|\partial
_{u}U_{\mathrm{sp}}(u)|}\gamma _{1}^{2}\left[ \varepsilon
_{1}w_{0}I_{0}(w_{0})I_{1}(w_{1})-\varepsilon
_{0}w_{1}I_{1}(w_{0})I_{0}(w_{1})\right] _{u=u_{\mathrm{sp}}}.  \label{Es}
\end{equation}%
\newline
Inside (outside) the cylinder, the radial dependence of the electric and
magnetic fields corresponding to the radiated surface polaritons are given
in terms of the function $I_{1}(\gamma _{0}u_{\mathrm{sp}}r/r_{c})$ ($%
K_{1}(\gamma _{1}u_{\mathrm{sp}}r/r_{c})$) for the components $E_{1}$ and $%
H_{2}$ and in terms of the function $I_{0}(\gamma _{0}u_{\mathrm{sp}%
}r/r_{c}) $ ($K_{0}(\gamma _{1}u_{\mathrm{sp}}r/r_{c})$) for $E_{3}$. The
latter is the only nonzero component on the axis of the cylinder.

We have considered an idealized linear annular beam. The corresponding
results describe the radiation intensity in the spectral range where the
radiation wavelength is much larger than the beam transverse and
longitudinal sizes. The corresponding results for the spectral density of
the radiation intensity can be generalized for a finite size annular beam
with azimuthally symmetric charge distribution $\rho (r,z)$. The Fourier
components of the fields corresponding to the annular element of the beam
with the charge $2\pi \rho (r^{\prime },z^{\prime })r^{\prime }dr^{\prime
}dz^{\prime }$ are obtained from the expressions given in Section \ref%
{sec:Fields} by the replacement%
\begin{equation}
q\rightarrow \rho (r^{\prime },z^{\prime })e^{-ik_{z}z^{\prime }}r^{\prime
}dr^{\prime }dz^{\prime },\;r_{0}\rightarrow r^{\prime }.  \label{qRepl}
\end{equation}%
The fields are obtained by the inverse Fourier transformation. For example,
denoting by $\mathbf{E}_{(\rho )}(x)$ the electric field generated by the
beam under consideration, we get%
\begin{equation}
\mathbf{E}_{(\rho )}(x)=\int_{0}^{\infty }dr^{\prime }\int_{-\infty
}^{+\infty }dz^{\prime }\,r^{\prime }\rho (r^{\prime },z^{\prime
})\int_{-\infty }^{\infty }dk_{z}\,e^{ik_{z}\left( z-z^{\prime }-vt\right) }%
\frac{\mathbf{E}_{k_{z}}(r)|_{r_{0}=r^{\prime }}}{q},  \label{Ero}
\end{equation}%
where the nonzero components of $\mathbf{E}_{k_{z}}(r)$ are given by (\ref%
{E13ex}) and (\ref{E13i}). The energy losses per unit length are expressed as%
\begin{equation}
\frac{dW_{(\rho )}}{dz}=2\pi \int_{0}^{\infty }dr\int_{-\infty }^{+\infty
}dz\,r\rho (r,z)E_{(\rho )3}(x).  \label{Wro}
\end{equation}

The radius of electron annular beams can be controlled by electric and
magnetic fields in a manner analogous to their formation (electromagnetic
lenses, see, e.g., \cite{AnBe05}). These fields can be used to separate the
beam and surface polaritons on the cylinder surface. Another method could be
placing an annular aperture blocking the beam and allowing the surface
polaritons continue to propagate. The surface polaritons can also be
separated by using reflecting structures at the end of cylindrical
waveguide. The reflection takes place also from the edge $z=z_{0}$ of the
finite length waveguide. Alternatively, one can use cylinders made of two
distinct materials with $\varepsilon _{0}^{\prime }<0$ in the region $z<z_{0}
$ and $\varepsilon _{0}^{\prime }>0$ in the region $z>z_{0}$. The surface
polaritons are not allowed to propagate in the region $z>z_{0}$ and they are
reflected back to the region $z<z_{0}$.

\section{Conclusion}

\label{sec:Conc}

In this study, we have examined the radiation of surface polaritons from an
annular beam of charged particles enclosing a cylindrical waveguide embedded
in a homogeneous medium. The electric and magnetic fields have been found by
using the Green dyadic for the geometry under consideration. The Fourier
components of the electric field outside and inside the cylinder are given
by the expressions (\ref{E13ex}) and (\ref{E13i}). The magnetic field is
directed along the azimuthal direction and the corresponding Fourier
component is connected to the electric field by the relation (\ref{H2ei}).
In the general case of dispersion for dielectric permittivities $\varepsilon
_{j}=\varepsilon _{j}(\omega )$, the spectral density of the energy losses
per unit time is given by (\ref{dE}), where the first term in the right-hand
side corresponds to the energy losses in a homogeneous medium with
permittivity $\varepsilon _{1}$. Depending on the spectral range and
dispersion law, the formula (\ref{dE}) describes different types of
radiation processes: Cherenkov radiation propagating outside the cylinder,
radiation on guiding modes of cylindrical waveguide and emission of surface
polaritons.

The surface polaritons are radiated in the spectral range where the real
parts of dielectric permittivities of the cylinder and surrounding medium
have opposite signs. The detailed consideration is presented for the case $%
\varepsilon _{0}^{\prime }<0<\varepsilon _{1}^{\prime }$. In this case the
general formula is specified to (\ref{dEsp}). For small values of the
imaginary part of the dielectric permittivity the spectral density of the
energy losses have strong narrow peaks centered at the frequencies
corresponding to the surface polaritonic eigenmodes of the dielectric
cylinder. They are roots of the equation (\ref{SPmodes}). The highest peaks
are obtained for intermediate values of the beam velocity. We have presented
the numerical results in the case of the Drude dispersion law for the
cylinder dielectric permittivity. The spectral density of the radiation
intensity for surface polaritons is displayed in Figs. \ref{fig2} and \ref%
{fig3} as a function of the radiation frequency in units of the plasma
frequency.

As the damping frequency decreases, the height of the peaks in the spectral
distribution of radiation intensity for surface polaritons increases, and
their width decreases. In the limit $\gamma \rightarrow 0$ the spectrum of
the surface polaritons becomes discrete with the eigenfrequencies determined
by the solutions of (\ref{SPmodes}). We have analytically demonstrated that
transition by using the relation (\ref{deltalim}). For a given $\beta _{1}$,
the surface polaritons are radiated under the condition (\ref{regroot}) for
the ratio of dielectric permittivities. The wavelength of the radiated
surface polaritons increases with increasing ratio $\varepsilon
_{1}/\varepsilon _{0}$ (see Fig. \ref{fig4}).

\section*{Acknowledgement}

A.A.S. was supported by the Science Committee of RA, in the frames of the
research project No. 21AG-1C047. L.Sh.G., H.F.K., and V.Kh.K were supported
by the Science Committee of RA, in the frames of the research project No.
21AG-1C069.


\begin{thebibliography}{99}
\bibitem{Maie07} S.A. Maier, Plasmonics: Fundamentals and Applications,
Springer, 2007.

\bibitem{Enoc12} S. Enoch, N. Bonod (Editors), Plasmonics: From Basics to
Advanced Topics, Springer, 2012.

\bibitem{Stoc18} M.I. Stockman \textit{et al.}, Roadmap on plasmonics, J.
Optics 20 (2018) 043001.

\bibitem{Marq08} R. Marqu\'{e}s, F. Mart\'{\i}n, M. Sorolla, Metamaterials
with Negative Parameters: Theory, Design, and Microwave Applications, John
Wiley \& Sons, Hoboken, NJ, 2008.

\bibitem{Han13} Zh. Han, S.I. Bozhevolnyi, Radiation guiding with surface
plasmon polaritons, Rep. Prog. Phys. 76 (2013) 016402.

\bibitem{Bolt11} A. Boltasseva, H.A. Atwater, Low-loss plasmonic
metamaterials, Science 331(6015) (2011) 290.

\bibitem{Wije15} T.M. Wijesinghe, M. Premaratne, G.P. Agrawal, Low-loss
dielectric-loaded graphene surface plasmon polariton waveguide based
biochemical sensor, J. Appl. Phys. 117 (2015) 213105.

\bibitem{Gonc16} P.A.D. Gon\c{c}alves, N.M.R. Peres, An Introduction to
Graphene Plasmonics, World Scientific, Singapore, 2016.

\bibitem{Haja18} Y. Hajati, Z. Zanbouri, M. Sabaeian, Low-loss and
high-performance mid-infrared plasmon-phonon in graphene-hexagonal boron
nitride waveguide, J. Opt. Soc. Am. B. 35 (2018) 446.

\bibitem{Zhen20} K. Zheng, Y. Yuan, L. Zhao, Y. Chen, F. Zhang, J. Song, J.
Qu, Ultra-compact, low-loss terahertz waveguide based on graphene plasmonic
technology, 2D Mater. 7 (2020) 015016.

\bibitem{Teng22} D. Teng, Z. Wang, Q. Huan, H. Wang, and K. Wang, A low loss
platform for subwavelength terahertz graphene plasmon propagation, Optical
Materials 128 (2022) 112436.

\bibitem{Qin23} X. Qin, Y. He, W. Sun, P. Fu, S. Wang, Z. Zhou, Y. Li,
Stepped waveguide metamaterials as low-loss effective replica of surface
plasmon polaritons, Nanophotonics 12(7) (2023) 1285. 

\bibitem{Bash06} M.V. Bashevoy, F. Jonsson, A.V. Krasavin, N.I. Zheludev, Y.
Chen, M.I. Stockman, Generation of traveling surface plasmon waves by
free-electron impact, Nano Lett. 6 (2006) 1113. 

\bibitem{Cai09} W. Cai, R. Sainidou, J. Xu, A. Polman, F.J. Garc\'{\i}a de
Abajo, Efficient generation of propagating plasmons by electron beams, Nano
Lett. 9 (2009) 1176. 

\bibitem{Liu12} S. Liu, P. Zhang, W. Liu, S. Gong, R. Zhong, Y. Zhang, M.
Hu, Surface polariton Cherenkov light radiation source, Phys. Rev. Lett. 109
(2012) 153902. 

\bibitem{Gong14} S. Gong, M. Hu, R. Zhong, X. Chen, P. Zhang, T. Zhao, S.
Liu, Electron beam excitation of surface plasmon polaritons, Opt. Express 22
(2014) 19252. 

\bibitem{Kuma16} P. Kumar, R. Kumar, S. Kumar Rajouria, Cherenkov terahertz
surface plasmon excitation by an electron beam over an ultrathin metal film,
J. Appl. Phys. 120 (2016) 223101. 

\bibitem{Gong17} S. Gong, M. Hu, R. Zhong, T. Zhao, Ch. Zhang, Sh. Liu,
Mediated coupling of surface plasmon polaritons by a moving electron beam,
Opt. Express 25 (2017) 25919. 

\bibitem{Zhan24} P. Zhang, Y. Dong, X. Li, X. Cao, Y. Yang, G. Yu, Sh. Yang,
Sh. Wang, Y. Gong, In-plane radiation of surface plasmon polaritons excited
by free electrons, Micromachines 15 (2024) 723.

\bibitem{Abaj10} F.J. Garc\'{\i}a de Abajo, Optical excitations in electron
microscopy, Rev. Mod. Phys. 82 (2010) 209. 

\bibitem{Jian17} Z. Jiang, D. Gu, M. Zhao, Q. Gu, Application of surface
plasmon polaritons on charged particle beam diagnostics, J. Phys.: Conf.
Series 1067 (2018) 072017. 

\bibitem{Riva00} A. Rivacoba, N. Zabala, J. Aizpurua, Image potential in
scanning transmission electron microscopy, Prog. Surf. Sci. 65 (2000) 1.

\bibitem{Kota18} A.S. Kotanjyan, A.R. Mkrtchyan, A.A. Saharian, V.Kh.
Kotanjyan, Radiation of surface waves from a charge rotating around a
dielectric cylinder, JINST 13 (2018) C01016.

\bibitem{Kota19} A.S. Kotanjyan, A.R. Mkrtchyan, A.A. Saharian, V.Kh.
Kotanjyan, Generation of surface polaritons in dielectric cylindrical
waveguides, Phys. Rev. Spec. Top. Accel. Beams 22 (2019) 040701.

\bibitem{Saha20} A.A. Saharian, L.Sh. Grigoryan, A.Kh. Grigorian, H.F.
Khachatryan, A.S. Kotanjyan, Cherenkov radiation and emission of surface
polaritons from charges moving paraxially outside a dielectric cylindrical
waveguide, Phys. Rev. A 102 (2020) 063517.

\bibitem{Saha23} A.A. Saharian, L.Sh. Grigoryan, A.S. Kotanjyan, H.F.
Khachatryan, Surface polariton excitation and energy losses by a charged
particle in cylindrical waveguides, Phys. Rev. A 107 (2023) 063513.

\bibitem{AnBe05} Annular Electron Beams, in: Pulsed Power. Springer, Boston,
MA, 2005, Chapter 23, pp. 413-432. https://doi.org/10.1007/0-306-48654-7\_23.

\bibitem{Lloy17} S.M. Lloyd, M. Babiker, G. Thirunavukkarasu, J. Yuan,
Electron vortices: Beams with orbital angular momentum, Rev. Mod. Phys. 89
(2017) 035004.

\bibitem{Liu24} Y. Liu, D. Wu, T. Liang, Zh. Sheng, X. He, Manipulation of
annular electron beams in plasmas, arXiv:2410.10196.

\bibitem{Grig95} L.Sh. Grigoryan, A.S. Kotanjyan, A.A. Saharian, Green
function of an electromagnetic field in cylindrically symmetric
inhomogeneous medium, Izv. Nats. Akad. Nauk Arm., Fiz. 30 (1995) 239 (Engl.
Transl.: J. Contemp. Phys.).

\bibitem{Saha24b} A.A. Saharian, S.B. Dabagov, H.F. Khachatryan, L.Sh.
Grigoryan, Quasidiscrete spectrum Cherenkov radiation by a charge moving
inside a dielectric waveguide, JINST 19 (2024) C06017.

\bibitem{Ashl74} J.C. Ashley, L.C. Emerson, Dispersion relations for
nonradiative surface plasmons on cylinders, Surf. Sci. 41 (1974) 615.

\bibitem{Khos91} H. Khosravi, D. R. Tilley, R. Loudon, Surface polaritons in
cylindrical optical fibers, J. Opt. Soc. Am. A 8 (1991) 112.
\end{thebibliography}
\end{document}